\documentclass[lettersize,journal]{IEEEtran}
\usepackage{amsmath,amsfonts}
\usepackage{algorithmic}
\usepackage{algorithm}
\usepackage{array}
\usepackage[caption=false,font=footnotesize,labelfont=sf,textfont=sf]{subfig}
\usepackage{textcomp}
\usepackage{stfloats}
\usepackage{url}
\usepackage[hidelinks]{hyperref}
\usepackage{xurl}
\usepackage{verbatim}
\usepackage{graphicx}
\usepackage{cite}
\usepackage[table]{xcolor}
\usepackage{booktabs}
\usepackage[normalem]{ulem}
\usepackage{multirow}
\usepackage{enumitem}
\newlist{tabitemize}{itemize}{1}
\setlist[tabitemize]{leftmargin=1.1em, labelsep=0.4em, nosep, label=\textbullet, before=\vspace{-5pt}}

\def\BibTeX{{\rm B\kern-.05em{\sc i\kern-.025em b}\kern-.08em
    T\kern-.1667em\lower.7ex\hbox{E}\kern-.125emX}}
\makeatletter
\def\subsubsection{\@startsection{subsubsection}{3}{1.25\parindent}{0.1ex plus 0.1ex minus 0.1ex}%
{0.1ex}{\normalfont\normalsize\itshape}}%
\makeatother

\begin{document}

\title{Large Language Models over Networks: Collaborative Intelligence under Resource Constraints}

\author{Liangqi Yuan,~\IEEEmembership{Graduate Student Member,~IEEE,}
    Wenzhi Fang,~\IEEEmembership{Graduate Student Member,~IEEE,}
    Shiqiang Wang,~\IEEEmembership{Fellow,~IEEE,}
    H. Vincent Poor,~\IEEEmembership{Life Fellow,~IEEE,}
    and Christopher G. Brinton,~\IEEEmembership{Senior Member,~IEEE}
        
\thanks{This project was supported in part by the National Science Foundation (NSF) under grants CNS-2146171 and CPS-2313109, the Office of Naval Research (ONR) under grant N00014-22-1-2305, and the Air Force Office of Scientific Research (AFOSR) under grant FA9550-24-1-0083.}
\thanks{L. Yuan, W. Fang, and C. G. Brinton are with Purdue University, IN, USA. E-mail: \{liangqiy, fang375, cgb\}@purdue.edu.}
\thanks{S. Wang is with University of Exeter, UK. E-mail: shiqiang.wang@ieee.org.}
\thanks{H. V. Poor is with Princeton University, USA. E-mail: poor@princeton.edu.}
}

\maketitle

\begin{abstract}
Large language models (LLMs) are transforming society, powering applications from smartphone assistants to autonomous driving. Yet cloud-based LLM services alone cannot serve a growing class of applications, including those operating under intermittent connectivity, sub-second latency budgets, data-residency constraints, or sustained high-volume inference. On-device deployment is in turn constrained by limited computation and memory. No single endpoint can deliver high-quality service across this spectrum. This article focuses on collaborative intelligence, a paradigm in which multiple independent LLMs distributed across device and cloud endpoints collaborate at the task level through natural language or structured messages. Such collaboration strives for superior response quality under heterogeneous resource constraints spanning computation, memory, communication, and cost across network tiers. We present collaborative inference along two complementary and composable dimensions: vertical device-cloud collaboration and horizontal multi-agent collaboration, which can be combined into hybrid topologies in practice. We then examine learning to collaborate, addressing the training of routing policies and the development of cooperative capabilities among LLMs. Finally, we identify open research challenges including scaling under resource heterogeneity and trustworthy collaborative intelligence.
\end{abstract}

\section{Introduction}

Large language models (LLMs) have shown remarkable capabilities in question answering, code generation, and multi-modal reasoning, powering applications from smartphone assistants to autonomous vehicles and robots. However, these capabilities rest on substantial computational and memory foundations. Frontier LLMs contain hundreds of billions to trillions of parameters, and both training and inference depend on resource-rich cloud data centers. At the same time, a growing number of applications cannot rely on cloud APIs alone. UAVs and field robots may enter connectivity-denied environments, closed-loop control and real-time agents cannot tolerate cloud round-trip latency, regulated domains such as healthcare and finance prohibit sensitive data from leaving the device, and sustained agentic workloads are bounded by per-token pricing and provider rate limits~\cite{zheng2025review}. Lightweight LLMs with up to a few billion parameters make on-device execution feasible but still lag frontier cloud LLMs on complex tasks.

Prior work on deploying LLMs across network tiers has largely operated at the model level. On the training side, federated learning enables collaborative fine-tuning through gradient or adapter exchange~\cite{cheng2025toward}. On the inference side, techniques such as model partitioning, speculative decoding, and context compression reduce the latency or communication cost of serving a single LLM across devices and servers~\cite{lin2025pushing}. However, these model-level approaches share common limitations, as they require white-box access to model internals and assume tightly coupled execution across endpoints. More fundamentally, they accelerate a single model rather than orchestrating cooperation among heterogeneous endpoints, including proprietary cloud APIs accessible only through text interfaces, that must jointly handle tasks no single endpoint can complete alone.

\begin{figure*}[t]
    \centering
    \includegraphics[width=\linewidth]{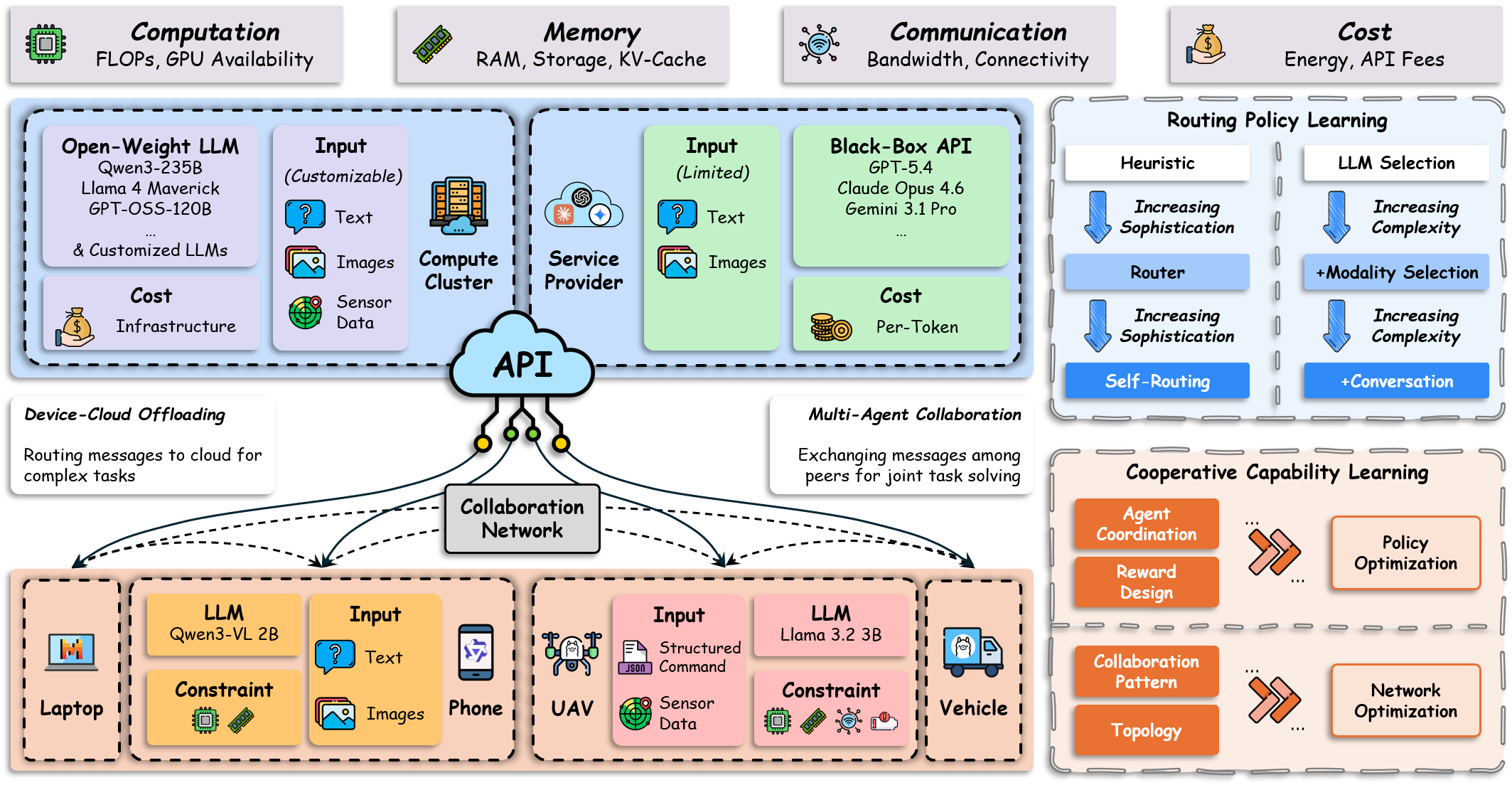}
    \vspace{-10pt}
    \caption{Collaborative LLM networks: device-cloud offloading and multi-agent collaboration enable heterogeneous LLM endpoints to jointly deliver higher response quality while navigating resource constraints across computation, memory, communication, and cost.}
    \vspace{-10pt}
    \label{fig:overview}
\end{figure*}

This article focuses on a distinct and increasingly important paradigm that we term \textbf{collaborative intelligence}, wherein multiple independent LLMs distributed across device and cloud endpoints collaborate at the task level through \textit{semantic exchanges of natural language or structured messages}. Each node runs a self-contained LLM and shares queries, responses, task descriptions, and observations rather than weights or intermediate tensors, as when smartphones offload complex queries to cloud LLMs, vehicles coordinate with home devices through structured commands, and cloud LLMs orchestrate UAV inspections by exchanging task descriptions with on-device LLMs (Fig.~\ref{fig:overview}). This design accommodates heterogeneous architectures without matched layer configurations, admits black-box API endpoints, and is supported by emerging standards such as the Agent-to-Agent (A2A) protocol. Such standards, however, specify only how endpoints communicate and are agnostic to resource constraints, leaving open whether to collaborate at all, which endpoint to involve, what context to keep, and what to transmit and when. Collaborative intelligence operates at the decision layer above them, resolving these questions under latency, communication, and cost budgets while preserving quality.

The remainder of this article is organized around five contributions. We introduce collaborative intelligence as a task-level paradigm for networked LLMs, motivated by a resource landscape spanning computation, memory, communication, and cost that makes such collaboration both necessary and difficult (Sec. \ref{sec:landscape}). Building on this, we present the system design of collaborative inference along two complementary and composable dimensions, vertical device-cloud and horizontal multi-agent (Sec. \ref{sec:architecture}). We then review the learning techniques that equip LLMs to route requests and cooperate effectively (Sec. \ref{sec:learning}). We further present two case studies, on device-cloud routing and on multi-agent coordination (Sec.~\ref{sec:case_study}). Finally, we identify open challenges in scaling under heterogeneity and in building trustworthy collaborative systems, and outline directions for future research (Sec. \ref{sec:open_challenges}).

\section{The Landscape of Networked LLMs: Challenges and Opportunities}
\label{sec:landscape}

\begin{figure*}[t]
    \centering
    \subfloat[On-Device LLM Benchmark\label{fig:device}]{\includegraphics[width=0.5\linewidth]{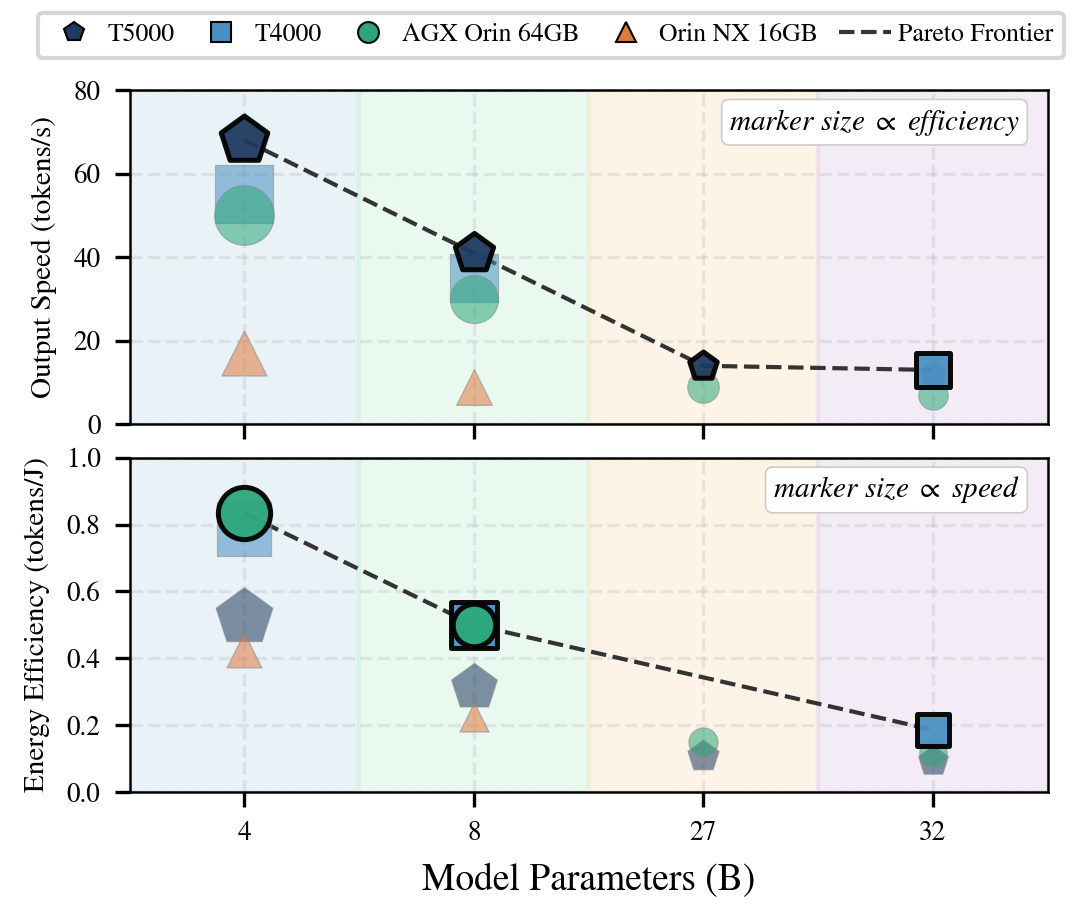}}
    \hfill
    \subfloat[Cloud LLM Benchmark\label{fig:cloud}]{\includegraphics[width=0.5\linewidth]{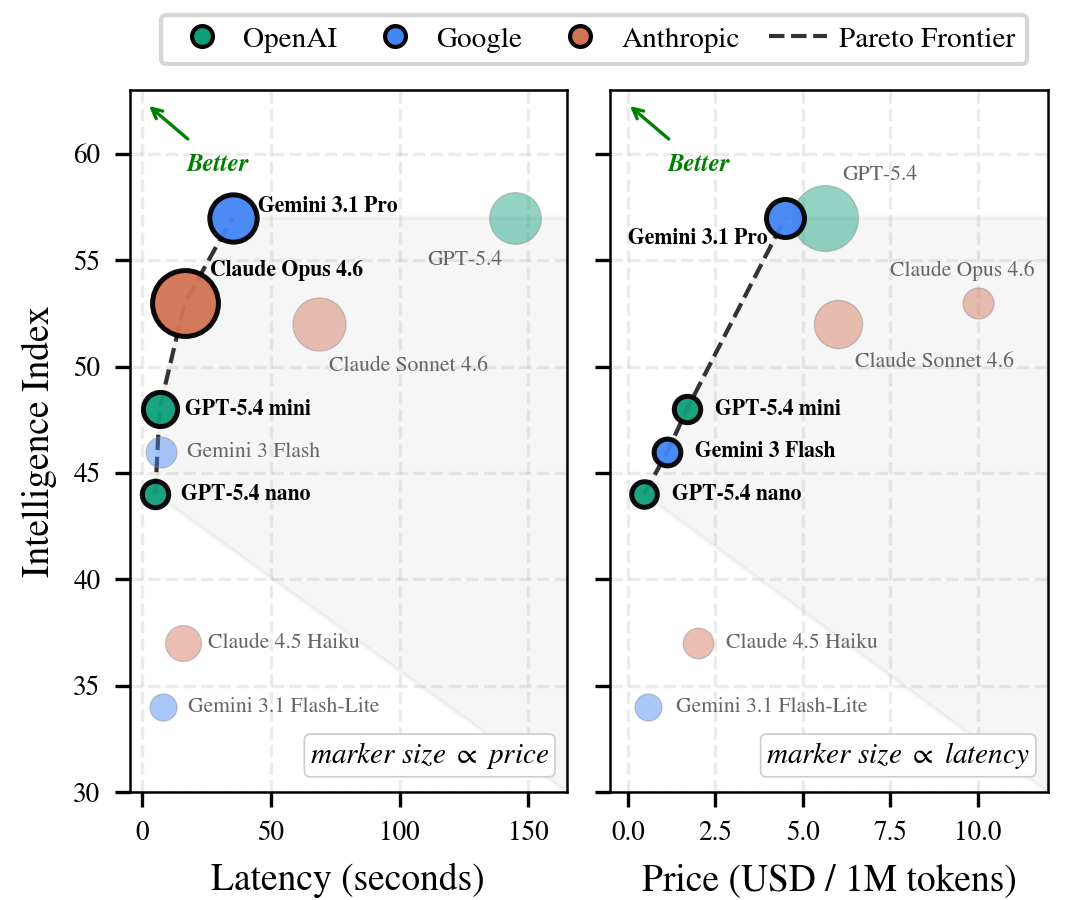}}
    \caption{Performance-resource tradeoffs across network tiers. The Pareto frontier shifts with the binding resource: in (a), the model leading on throughput does not lie on the energy-efficiency Pareto frontier. In (b), cloud models on the latency-quality frontier are no longer Pareto-optimal under price-quality, and vice versa. No single endpoint dominates across all dimensions.\protect\footnotemark}
    \vspace{-10pt}
    \label{fig:benchmarks}
\end{figure*}

Today's LLM ecosystem spans a wide resource spectrum, from lightweight on-device LLMs to cloud frontier LLMs. Table~\ref{tab:resource_constraints} groups the constraints across this spectrum into four categories, namely computation, memory, communication, and cost~\cite{qu2025mobile}. Navigating this spectrum means deciding which endpoint serves each request, a form of routing that is not new in itself, since cloud services already route queries among models to cut cost while preserving quality. But that router runs in the data center with global visibility and an abundance of compute, and its limits are commercial rather than physical. Networked LLMs face a different regime. Constraints are physical and fixed at runtime by on-device compute, memory, and battery and by intermittent, bandwidth-limited links. The candidate endpoints differ by orders of magnitude in capability, latency, and cost rather than being variants of one model. Communication itself becomes a decision variable, and data-residency rules can forbid offloading regardless of budget. The decision engine, moreover, runs on the constrained device under only partial information, so it is subject to the same limits it manages. These four categories are tightly coupled, since quantizing an LLM to fit device memory sacrifices quality while offloading to the cloud trades compute for latency and cost. As Fig.~\ref{fig:benchmarks} shows, no single LLM dominates every dimension, with significant heterogeneity existing both across and within tiers.

\begin{table*}[t]
    \caption{Resource Constraints in Networked LLM Environments and Triggered Collaboration Tradeoffs}
    \label{tab:resource_constraints}
    \centering
    \resizebox{\linewidth}{!}{%
    \begin{tabular}{@{}p{1.5cm}p{5.1cm}p{5.2cm}p{5.4cm}@{}}
    \toprule
    \textbf{Resource} & \textbf{On-Device} & \textbf{Cloud} & \textbf{Collaboration Tradeoff} \\
    \midrule
    \textbf{Computation}
    &
    \textit{FLOPs, GPU Availability}\par\vspace{5pt}
    \begin{tabitemize}
    \item Compute resources shared with other tasks
    \item High per-token latency
    \end{tabitemize}
    &
    \textit{GPU Clusters, Rate Limits}\par\vspace{5pt}
    \begin{tabitemize}
    \item Abundant but shared, subject to queuing
    \item Rate limits cap throughput
    \end{tabitemize}
    &
    \textit{Where to Run}\par\vspace{5pt}
    \begin{tabitemize}
    \item Offload heavy reasoning, keep light local
    \item Trade device energy for cloud cost
    \end{tabitemize}
    \\
    \midrule
    \textbf{Memory}
    &
    \textit{RAM, Storage, KV-Cache}\par\vspace{5pt}
    \begin{tabitemize}
    \item Constrained LLM sizes
    \item Restricted KV-cache limits context length
    \end{tabitemize}
    &
    \textit{VRAM, Context Window}\par\vspace{5pt}
    \begin{tabitemize}
    \item Memory allocation not user-configurable
    \item Context window capped by provider
    \end{tabitemize}
    &
    \textit{What Context to Keep}\par\vspace{5pt}
    \begin{tabitemize}
    \item Retain or transmit history and KV-cache
    \item Match model and context length to the task
    \end{tabitemize}
    \\
    \midrule
    \textbf{Communi- \newline cation}
    &
    \textit{Bandwidth, Connectivity}\par\vspace{5pt}
    \begin{tabitemize}
    \item High round-trip latency
    \item Intermittent and unstable wireless links
    \end{tabitemize}
    &
    \textit{Throughput, Latency}\par\vspace{5pt}
    \begin{tabitemize}
    \item Throttled responses under rate limits
    \item Streaming delivery depends on link quality
    \end{tabitemize}
    &
    \textit{What and When to Transmit}\par\vspace{5pt}
    \begin{tabitemize}
    \item Select modalities and context volume to send
    \item Switch endpoints only when the link permits
    \end{tabitemize}
    \\
    \midrule
    \textbf{Cost}
    &
    \textit{Energy, Hardware}\par\vspace{5pt}
    \begin{tabitemize}
    \item Finite battery limits sustained inference
    \item Expense for device compute hardware
    \end{tabitemize}
    &
    \textit{Infrastructure, API Fees}\par\vspace{5pt}
    \begin{tabitemize}
    \item Infrastructure fees for cloud compute
    \item Per-token pricing bounds total queries
    \end{tabitemize}
    &
    \textit{How to Spend the Budget}\par\vspace{5pt}
    \begin{tabitemize}
    \item Allocate latency and cost budget across turns
    \item Offload only when the cloud call pays off
    \end{tabitemize}
    \\
    \bottomrule
    \end{tabular}
    }
\end{table*}

The binding constraints vary significantly from one device to another, as illustrated in Fig.~\ref{fig:overview}. A smartphone running a quantized LLM is bottlenecked primarily by on-device computation and memory, since it sustains only moderate token throughput and must aggressively quantize the model to fit within a few gigabytes of RAM, directly limiting local inference quality and context length. A UAV faces a fundamentally different profile, with all resources scarce simultaneously. Limited onboard compute and memory are compounded by intermittent wireless connectivity and a strict energy budget tied to battery capacity and flight time. For such devices, selectively offloading to the cloud can be attractive when connectivity permits, since local inference is itself expensive in both latency and power. Yet the decision must also weigh link reliability and the privacy sensitivity of the transmitted data, which often preclude cloud offloading in mission-critical scenarios. Input modalities differ just as widely, from text and photos on a smartphone to aerial imagery and flight telemetry on a UAV, ruling out any one-size-fits-all strategy.

These observations point to both an opportunity and a set of challenges. Because device and cloud LLMs have complementary strengths, enabling them to collaborate at the task level can deliver a quality-latency-cost balance beyond what any individual endpoint offers. Yet realizing such collaboration is far from straightforward. Routing decisions must be made automatically and continuously, jointly navigating multiple competing objectives rather than optimizing any single one. Multi-turn and agentic workflows compound the difficulty, requiring dialogue history and intermediate observations to be handed off across endpoints. Horizontal cooperation among peer devices adds further design choices around communication topology and message format, which interact with response quality in non-obvious ways. The remaining sections describe how architectural designs and learning techniques address these challenges.

\footnotetext{Device benchmarks from \url{https://www.jetson-ai-lab.com/models/} (accessed Apr. 2026), and cloud benchmarks from \url{https://artificialanalysis.ai/leaderboards/models} (accessed Apr. 2026).}

\section{Collaborative Inference Architecture}
\label{sec:architecture}

Collaborative inference can be organized along two complementary topologies, illustrated in Fig.~\ref{fig:overview}: vertical device-cloud collaboration, where on-device LLMs offload difficult tasks upward to more capable cloud LLMs, and horizontal multi-agent collaboration, where peer LLM agents collaborate to solve tasks collectively. The two are not mutually exclusive and are often composed into hybrid topologies in practice, as when a cloud LLM coordinates a fleet of on-device agents that also exchange messages directly with one another. Beyond the choice of topology, collaborative intelligence is defined by a layer of resource-aware decisions, namely whether to offload, which endpoint to use, and what to transmit under latency, bandwidth, energy, and cost budgets. Because collaboration always costs latency, energy, or money while its benefit saturates, these decisions reduce to invoking collaboration only where its marginal gain outweighs its cost.

\subsection{Device-Cloud Collaboration}

Device-cloud collaboration lets an on-device LLM answer what it can and escalate the rest to the cloud. The challenge is that routing happens before a request's complexity is known, and the endpoints differ not only in latency and cost but in answer quality. Routing is a bet on latent complexity, and because context accumulates on whichever endpoint serves a turn, each bet persists into the next.

\textbf{Complexity- and Resource-Aware LLM Selection.} For each request the system selects which LLM serves it and how much of the input to transmit. Both decisions depend on a quantity that is not directly observable, namely the complexity of the request. Complexity is semantic rather than structural, so surface features such as input length correlate only weakly with whether the on-device LLM can succeed, and multi-modal inputs widen this gap because an added image or document may prove either decisive or redundant depending on the request~\cite{yuan2025local}. The routing decision therefore rests on the engine's own estimate of request complexity, obtained for instance from the on-device model's response to the request, rather than on fixed rules over surface features. This estimate is then weighed against the supply side, where candidate endpoints differ in capability, latency, and monetary cost and these properties fluctuate with network conditions and server load, so the same request may warrant escalation over a fast link yet local handling over a congested one. Optimizing endpoint and payload jointly against these signals reduces communication cost while preserving response quality.

\textbf{Session-Level Routing under Accumulating Context.} In multi-turn conversations and agentic workflows, the endpoint chosen for one turn shapes the cost of the next. Serving consecutive turns on the same endpoint reuses the KV cache already held there, so only the new tokens are prefilled, whereas switching endpoints requires transmitting the prior dialogue and re-prefilling the whole history to rebuild that cache. From a resource standpoint this favors staying put, as both the transferred history and the recomputed cache grow with conversation length. From a complexity standpoint the opposite holds, since turns within a session vary and the endpoint suited to a routine turn is seldom the one a later reasoning-heavy turn warrants. The tension between the two makes routing a sequential problem rather than a per-turn one. Because the switching cost rises with accumulated context, the incentive to move to a better-matched endpoint erodes as the conversation lengthens, so greedy per-turn routing gradually locks a session onto its incumbent endpoint even when later turns would be served better elsewhere. Context compression can lower this switching cost but is lossy and leaves the coupling intact~\cite{jiang2023llmlingua}, so effective routing looks ahead across the session, weighing a turn's complexity against the complexity likely to follow, rather than optimizing each turn in isolation.

\subsection{Multi-Agent Collaboration}

Horizontal collaboration runs multiple LLM agents as peers that coordinate by exchanging messages, rather than deferring to a stronger agent as in device-cloud collaboration. It helps by decomposing a task into parallel subtasks or by combining diverse perspectives on the same task, and the agents may reside on devices, in the cloud, or across both tiers.

\textbf{Collaboration Patterns From Datacenter to Network.} Multi-agent collaboration can be characterized by interaction pattern (e.g., debate, division-of-labor, hierarchical), execution structure (e.g., parallel, sequential), and coordination scope (e.g., centralized, decentralized), each with different implications for response quality, latency, communication, and cost~\cite{luo2025toward}, as illustrated in Fig.~\ref{fig:overview}. These patterns are usually characterized for agents inside a datacenter, and running them over a network turns six assumptions the datacenter grants for free into explicit design choices: (i) reliability, since messages may be lost; (ii) synchrony, since without a global clock exchanges are asynchronous rather than lockstep rounds and agreement is no longer guaranteed; (iii) shared state, since each agent holds a local view and reaches only nearby peers; (iv) free communication, since every message now costs latency, energy, and money; (v) stable membership, since agents may join, leave, or partition mid-task; and (vi) homogeneity, since heterogeneous agents may interpret the same message differently. These degrees of freedom favor collaboration that is asynchronous, locality-aware, and robust to partial or outdated information, and they rank the patterns, since debate leans hardest on the datacenter assumptions through its repeated, synchronous all-to-all critique rounds, whereas division-of-labor and hierarchical coordination ask less of the network and port more readily.

\textbf{Communication Topology and Overhead.} Communication cost here is set by the topology and, because each message is text rather than a compact numerical vector, by message length~\cite{zhang2025cut}. A fully connected graph scales quadratically in messages, while star and tree structures cut or centralize this load at some loss of sharing. Debate is the most exposed, since $N$ agents over $T$ rounds exchange $O(N^2)$ messages per round while each lengthens with the exchanges it carries, giving $O(N^2 T^2)$ traffic in the worst case, whereas hierarchical coordination is far cheaper as only the supervisor talks to all workers. This cost ranking mirrors the robustness ranking above, so debate pays twice for the network, yet denser connectivity does not reliably improve the answer, as Sec.~\ref{sec:learning_cooperative} discusses. Because the best topology depends on bandwidth and task conditions that shift across a network, fixing it once is rarely right, and text-level exchange makes correction cheap, since rewiring reroutes messages with no model state to migrate as in model-parallel execution. The graph can thus thin at runtime, and a node that drops out takes only its own edges with it. As total traffic scales with the number of active edges times message length, the two are substitutable levers, so the same quality can be sustained at very different costs by sparsifying the graph when bandwidth binds and compressing messages when per-token price does.

\section{Learning to Collaborate}
\label{sec:learning}

The preceding section described how LLMs collaborate. This section turns to how those collaboration strategies are learned, with two objectives that differ in what is learned. Learning to route learns a selection policy that decides which endpoint should answer each request. Learning to cooperate learns interaction itself, jointly training agents to complement one another. The right panel of Fig.~\ref{fig:overview} summarizes this landscape.

\subsection{Routing Policy Learning}

In device-cloud collaboration, every incoming request requires a routing decision between local and cloud processing. Rather than relying on hand-crafted rules, recent work trains routing policies that learn to make this choice by balancing quality, latency, and cost.

\textbf{Router-Based Selection.} A natural starting point is a lightweight classifier that examines each incoming request and routes it to the most suitable LLM~\cite{ding2024hybrid}. Candidate LLMs are first profiled offline for quality, latency, and per-token cost on representative benchmarks, and a compact classifier (e.g., a fine-tuned BERT model) then learns to predict which candidate offers the best tradeoff for each request, typically by maximizing expected quality under a budget on average cost or latency. Such classifiers treat each request independently, whereas in multi-turn conversations switching between local and cloud LLMs mid-session requires transferring the accumulated dialogue context, which couples the routing decisions across turns. The routing policy is then better modeled as a Markov decision process in which the state encodes the current request, conversation history, and resource usage so far, and the action selects an endpoint for the current turn. A small policy network separate from the frozen LLMs is trained by rolling out conversations, where each turn is forwarded to the selected endpoint as a query, the returned response is scored, and the reward combines a bonus for that score with a penalty for the latency and cost incurred, so that reinforcement learning optimizes the cumulative tradeoff over an entire conversation rather than a single request~\cite{yuan2025local}.

\textbf{Self-Routing via Post-Training.} An alternative eliminates the external router altogether: the on-device LLM is trained to attempt the task first and inspect its own reasoning before deciding whether to escalate~\cite{fang2025collaborative}. Concretely, the LLM is post-trained using reinforcement learning with a composite reward, where it receives a quality bonus when its local answer is correct and a cost penalty whenever it escalates to the cloud, incentivizing the LLM to handle as many queries locally as possible without sacrificing accuracy. At inference time, the on-device LLM can draw on chain-of-thought signals generated during its own reasoning process, such as self-consistency across sampled traces, entropy of intermediate steps, or explicit self-evaluation tokens, to gauge whether the current problem lies within its capability. Because the routing decision is made after the LLM has already engaged with the problem rather than from surface-level features of the prompt alone, self-routing yields substantially more accurate offloading than classifier-based routers do. The two families thus trade the cost of the routing signal against its accuracy, since an external router reads cheap surface features whereas self-routing pays for a local attempt to judge whether escalation is worth its price.

\subsection{Cooperative Capability Learning}
\label{sec:learning_cooperative}

LLMs are pre-trained in isolation and have never experienced genuine multi-party interaction, which limits what prompt engineering alone can achieve and motivates fine-tuning agents to collaborate explicitly.

\textbf{Cooperative Policy Optimization.} Current prompt-based systems assign roles and instructions but leave agents to improvise their interaction, often producing repetitive exchanges that fail to converge or even degrade over additional rounds. Training-based methods instead place multiple LLMs in collaborative scenarios and optimize their behavior through reinforcement learning under a shared task reward. A key finding is that training agents in isolation fails, since an agent optimized against a fixed partner learns to act independently rather than adapting to evolving strategies~\cite{park2025maporl}. Genuine cooperation emerges only when all participating agents are co-trained simultaneously. This co-training, however, presumes a fixed team trained together, whereas networked agents are heterogeneous and often owned by different parties, so the more pressing opportunity is cooperation that transfers to partners an agent was never trained with. Beyond answering offloaded queries, a capable cloud LLM can also coordinate device-side agents, deciding which to activate and when to conclude the discussion, with such policies learned through reinforcement learning~\cite{dang2025multi}.

\textbf{Inter-Agent Network Optimization.} Whereas cooperative training shapes the content of inter-agent messages, the network layer concerns which of those messages are necessary at all, and because LLM agents can reason, a substantial fraction is not. An agent can often infer a peer's likely contribution, or recover it from a shared observation, instead of eliciting it through an explicit exchange. Existing work curtails such redundant traffic by sparsifying the fully connected graph so that each agent observes only a subset of its peers~\cite{li2024improving}. A more fundamental lever, available only to reasoning agents, is to let observation and inference substitute for communication, incurring a message only when a peer can supply information the agent cannot derive independently. Topology optimization in this setting thus differs in kind from classical network design over passive relays, which allocates links solely to transport bits efficiently, since the connectivity worth maintaining is governed by what each agent knows, what it can infer, and its capability and resource budget, and evolves as the task unfolds and the agents' knowledge accumulates.

\section{Case Studies: From Routing to Coordination}
\label{sec:case_study}

The two case studies instantiate the tradeoffs of Sec.~\ref{sec:landscape} in settings that differ in task, resource constraints, and the roles assigned to device and cloud. The first is a multi-turn, multi-modal conversation constrained by per-session latency and usage cost, in which the cloud LLM serves as an escalation target for the on-device LLM, and a router determines, given the accumulated dialogue, which turns and which modalities are escalated. The second is a multi-agent coordination task constrained by a per-episode step budget shared between acting and communicating, together with a cap on cloud calls, in which the cloud LLM serves only as an advisor while on-device agents act as peers, and each agent determines at every step whether to act, message a peer, or consult the cloud.

\begin{figure}[t]
    \centering
    \includegraphics[width=\linewidth]{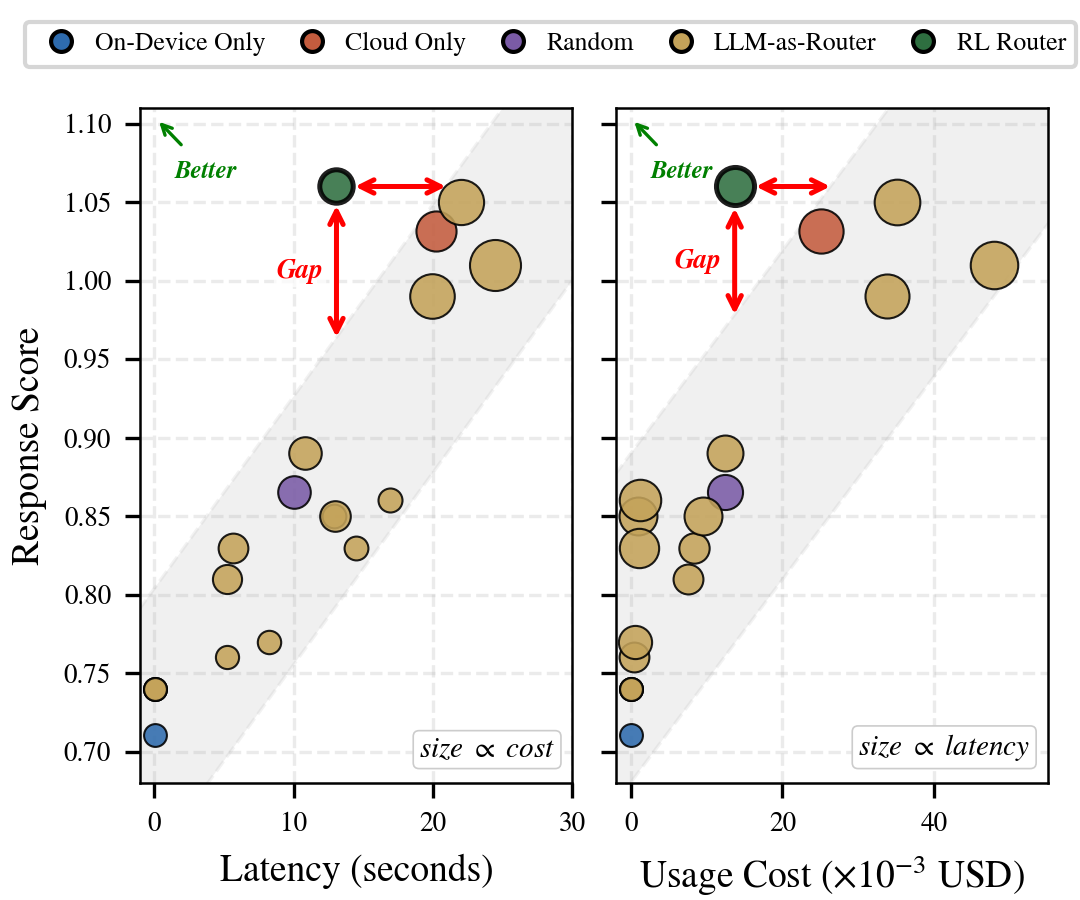}
    \vspace{-20pt}
    \caption{Quality-latency-cost tradeoffs for device-cloud routing. Baselines trace a common tradeoff band (shaded), while the RL router escapes it by carrying the remaining budget in its state rather than deciding per request.}
    \vspace{-10pt}
    \label{fig:TMO}
\end{figure}

\textbf{Device-Cloud Routing in Multi-Modal Conversations.} The first case study is a multi-turn, multi-modal conversation~\cite{yuan2025local} in which the appropriate endpoint shifts from turn to turn with the complexity of each request, from text-only edits that a small on-device LLM (Phi-3-mini) handles well to scene-grounded questions that need visual context and a frontier cloud LLM (GPT-4o). The visual context comes from partially redundant first-person, overhead, and side camera views, so which views to transmit is a second decision, as sending all of them wastes budget while sending too few withholds what a complex turn needs. As shown in Fig.~\ref{fig:TMO}, on-device-only inference is lowest in both cost and quality, cloud-only is high in quality but costly, and prompt-based LLM routers lie within the same band as the other baselines, in which higher quality comes only with higher latency and cost, since they decide where to run and what to transmit one request at a time. The learned RL policy departs from this band, matching cloud-only quality at 35\% lower latency and 45\% lower cost. The value of learned routing therefore lies less in per-request classification than in treating the resource budget as a stateful, temporal constraint.

\begin{figure}[t]
    \centering
    \includegraphics[width=\linewidth]{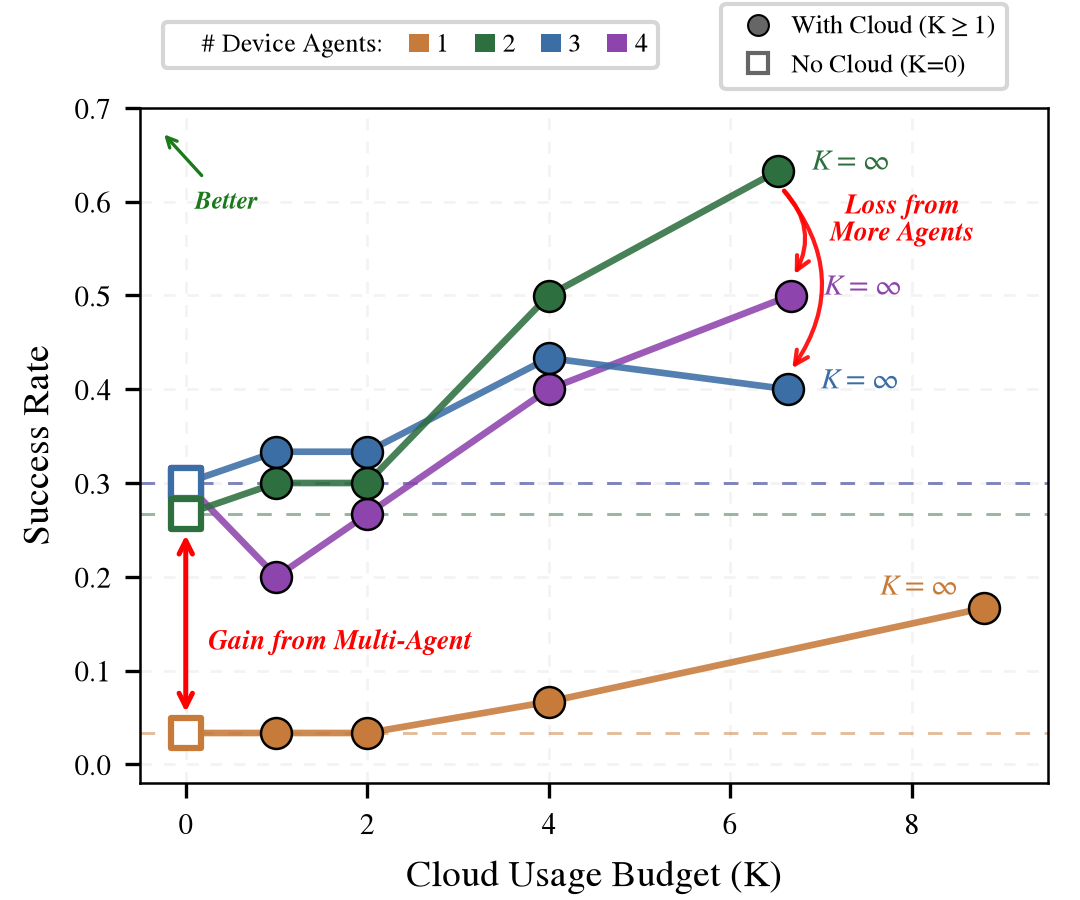}
    \vspace{-20pt}
    \caption{Multi-agent coordination with an on-demand cloud advisor. Peer collaboration drives most of the gain, yet success peaks at two agents at larger cloud budgets and declines beyond that point, as coordination overhead outweighs the added capacity.}
    \vspace{-10pt}
    \label{fig:overcooked}
\end{figure}

\textbf{Multi-Agent Coordination with On-Demand Cloud Calls.} The second case study is a multi-agent coordination task with interdependent subtasks, scored by task completion~\cite{sun2025collab}. On-device agents (Qwen3-4B) act synchronously over 50 steps per episode, each message or cloud call consuming one step. A cloud agent (Gemini 3 Flash) sees the global state but never acts, answering only when asked under a per-episode cloud usage budget $K$. As shown in Fig.~\ref{fig:overcooked}, a lone agent almost always fails, and adding a second agent lifts success from 3\% to 27\% even with no cloud, the largest gain from any added agent. Cloud calls cannot substitute for that second agent, as a lone agent with unlimited calls remains below two agents that never call the cloud. The two are complementary, as unlimited cloud calls raise success at every team size and more than double it for two agents, to 63\%. Beyond two agents, however, the gain reverses, since three or four agents underperform two at larger cloud budgets. Each added agent adds action capacity but must spend steps coordinating and contending over shared subtasks, and this overhead grows with team size. The budget decision thus shifts from whether to call the cloud to how many steps coordination may consume, and collaboration is beneficial only where the capacity gained outweighs that cost.

\section{Open Challenges and Research Directions}
\label{sec:open_challenges}

Deploying collaborative LLM systems over networks at scale exposes a range of unsolved problems. Below we highlight two broad axes of open challenges and identify promising directions for each.

\subsection{Scaling Under Resource Heterogeneity}
 
\textit{Adapting to constantly shifting ecosystems:} The resource spectrum that makes collaboration attractive also makes it difficult to engineer. A single collaboration framework must accommodate endpoints whose compute, memory, bandwidth, and cost budgets differ by orders of magnitude, as outlined in Table~\ref{tab:resource_constraints}, and these gaps will only widen as frontier cloud LLMs continue to grow while on-device deployment pushes toward ever smaller form factors. Today's solutions typically target a fixed set of LLMs and hardware profiles. An important open question is how to build collaboration protocols that remain effective as new LLMs and devices continuously enter and leave the ecosystem, without requiring retraining or manual reconfiguration each time.

\textit{From pairwise collaboration to large-scale swarms:} Most existing work studies collaboration among a handful of agents. Scaling to hundreds or thousands of LLM instances, such as a cloud coordinator managing a fleet of on-device agents in autonomous vehicles or drone swarms, raises challenges at every level. Scheduling must handle continuous streams of inference requests under strict latency budgets, agents must reach consensus over volatile and bandwidth-limited links, and individual nodes must reason autonomously during connectivity gaps. At the same time, many high-value tasks require maintaining coherent context across dozens of inference steps, and serving such long-context workloads concurrently for many agents imposes severe memory and communication pressure. How to co-design scheduling, communication, and fault-tolerance mechanisms that operate reliably across all these levels remains largely unexplored.

\subsection{Trustworthy Collaborative Intelligence}

\textit{Keeping sensitive data under control:} Collaborative inference moves queries across the device-cloud boundary, making it a trust boundary, not a compute split. Prompts may carry personal, medical, or financial data, and agentic LLMs with access to local files can leak such content into responses and propagate it across a multi-agent chain. This creates a tension between openness and containment, since more shared context improves quality but each piece transmitted is a potential exposure. Emerging approaches soften this tradeoff by keeping sensitive spans on the device and letting the cloud reason over typed placeholders that preserve each value's role while discarding its content~\cite{yuan2026paac}. Yet these designs presuppose that every sensitive span can first be identified in free-form text, which is context-dependent and unschematized, so detection is inexact and anything missed leaks in the clear, offering best-effort rather than provable privacy.

\textit{Defending against cascading attacks:} Prompt injection is well studied as a single-interface threat, but distributed LLM deployments introduce a new dimension, where a successful injection at any node can propagate along the collaboration hierarchy and contaminate downstream agents across the entire workflow. Injected instructions can persist covertly within accumulated context, interfering with subsequent inference steps long after the initial attack. The risk is amplified when LLM agents directly control physical systems such as vehicles or robots, where a compromised output can cause real-world harm. Understanding how adversarial inputs propagate through multi-tier collaborative systems, and developing defenses that can detect and contain such cascading effects in real time, are critical open problems.

\section{Conclusion}

As LLMs become embedded in every tier of the network, no single endpoint can meet the full spectrum of user demands on its own. Collaborative intelligence enables heterogeneous LLMs to cooperate at the task level through natural language and structured messages, achieving quality-latency-cost tradeoffs beyond any individual model. This article has organized the landscape along two complementary dimensions, device-cloud offloading and multi-agent collaboration, and examined how routing policies and cooperative training can make such collaboration effective. Realizing this vision at scale will require progress on several fronts, from protocols that adapt seamlessly as models and devices enter and leave the ecosystem, to trust and safety mechanisms that keep pace with the growing autonomy of LLM agents. We believe collaborative intelligence will become a foundational design principle for networked AI systems in the years ahead.

\section*{Acknowledgment}
The authors thank Dr. Dusit Niyato for his insightful feedback on this paper.

\bibliographystyle{IEEEtran}
\small\bibliography{reference}

\vspace{-0.4in}

\begin{IEEEbiographynophoto}
{Liangqi Yuan}
is pursuing the Ph.D. degree in ECE with Purdue University.
\end{IEEEbiographynophoto}

\vspace{-0.4in}

\begin{IEEEbiographynophoto}
{Wenzhi Fang}
is pursuing the Ph.D. degree in ECE with Purdue University.
\end{IEEEbiographynophoto}

\vspace{-0.4in}

\begin{IEEEbiographynophoto}
{Shiqiang Wang}
is a Professor of Artificial Intelligence in the Department of Computer Science, University of Exeter.
\end{IEEEbiographynophoto}

\vspace{-0.4in}

\begin{IEEEbiographynophoto}
{H. Vincent Poor}
is the Michael Henry Strater University Professor at Princeton University.
\end{IEEEbiographynophoto}

\vspace{-0.4in}

\begin{IEEEbiographynophoto}
{Christopher G. Brinton}
is the Elmore Associate Professor of ECE with Purdue University.
\end{IEEEbiographynophoto}

\end{document}